\documentclass[aps,prl,twocolumn,linenumber,floatfix,citeautoscript,superscriptaddress]{revtex4-1}
\usepackage{graphicx}
\usepackage{bm,amsmath,amssymb,mathrsfs,dcolumn}
\usepackage[colorlinks=true, linkcolor=blue, citecolor=blue, urlcolor=blue, linktoc=page, bookmarks=false, pdfstartview={FitH}, pdfborder={0 0 0.0 [3 3]}]{hyperref} % HyperLink
\usepackage[usenames]{color}
\hypersetup{pdfborder=0 0 0,colorlinks=true,citecolor=blue,linkcolor=blue}
\usepackage{epstopdf}
\usepackage{subfigure}
\usepackage{units}
\usepackage{lineno}
\usepackage{url}
\usepackage{color}
\usepackage{multirow}
\usepackage{graphicx}
\usepackage{booktabs}

\begin{document}
	
	\title{Reactor neutrino physics potentials of cryogenic pure-CsI crystal}
	
	\author{Lei Wang}
	\affiliation{Experimental Physics Division, Institute of High Energy Physics, Chinese Academy of Sciences, Beijing 100049, China}
	\affiliation{School of Physics, University of Chinese Academy of Sciences, Beijing, 100049, China}
	
	\author{Guanda Li}
	\affiliation{School of Nuclear Science and Engineering, North China Electric Power University, Beijing, 102206, China}
	
	\author{Zeyuan Yu}
	\affiliation{Experimental Physics Division, Institute of High Energy Physics, Chinese Academy of Sciences, Beijing 100049, China}
	\affiliation{School of Physics, University of Chinese Academy of Sciences, Beijing, 100049, China}
	\affiliation{State Key Laboratory of Particle Detection and Electronics, Beijing, 100049, China}
	
	\author{Xiaohua Liang}
	\affiliation{Astro-particle Physics Division, Institute of High Energy Physics, Chinese Academy of Science, Beijing, 100049, China}
	
	\author{Tian'an Wang}
	\affiliation{State Key Laboratory of High Power Semiconductor Laser, College of Physics, Changchun University of Science and Technology, Changchun, Jilin, 130022, China}
	
	\author{Fang Liu}
	\affiliation{Beijing Key Laboratory of Passive Safety Technology for Nuclear Energy, School of Nuclear Science and Engineering, North China Electric Power University, Beijing, 102206, China}

	\author{Xilei Sun}
	\altaffiliation{Corresponding author: sunxl@ihep.ac.cn}
	\affiliation{Experimental Physics Division, Institute of High Energy Physics, Chinese Academy of Sciences, Beijing, 100049, China}
	\affiliation{School of Physics, University of Chinese Academy of Sciences, Beijing, 100049, China}
	\affiliation{State Key Laboratory of Particle Detection and Electronics, Beijing, 100049, China}	
	
	\author{Cong Guo}
	\altaffiliation{Corresponding author: guocong@ihep.ac.cn}
	\affiliation{Experimental Physics Division, Institute of High Energy Physics, Chinese Academy of Sciences, Beijing 100049, China}
	\affiliation{School of Physics, University of Chinese Academy of Sciences, Beijing, 100049, China}
	\affiliation{State Key Laboratory of Particle Detection and Electronics, Beijing, 100049, China}
	
	\author{Xin Zhang}
	\affiliation{Experimental Physics Division, Institute of High Energy Physics, Chinese Academy of Sciences, Beijing 100049, China}
	\affiliation{School of Physics, University of Chinese Academy of Sciences, Beijing, 100049, China}
	
	\author{Yu Lei}
	\affiliation{ School of Electronic, Electrical Engineering and Physics, Fujian University of Technology, Fuzhou, 350118, China}

	\author{Yuede Chen}
	\affiliation{ School of Electronic, Electrical Engineering and Physics, Fujian University of Technology, Fuzhou, 350118, China}

	\date{\today }
	
	\begin{abstract}
	This paper presents a world-leading scintillation light yield among inorganic crystals measured from a 0.5~kg pure-CsI detector operated at 77 Kelvin. Scintillation photons were detected by two 2-inch Hamamatsu SiPM arrays equipped with cryogenic front-end electronics. Benefiting the light yield enhancement of pure-CsI at low temperatures and the high photon detection efficiency of SiPM, a light yield of 30.1 photoelectrons per keV energy deposit was obtained for X-rays and $\gamma$-rays with energies from 5.9~keV to 59.6~keV. Instrumental and physical effects in the light yield measurement are carefully analyzed. This is the first stable cryogenic operation of kg-scale pure-CsI crystal readout by SiPM arrays at liquid nitrogen temperatures for several days. The world-leading light yield opens a door for the usage of pure-CsI crystal in several fields, particularly in detecting the coherent elastic neutrino-nucleus scattering of reactor neutrinos. The potential of using pure-CsI crystals in neutrino physics is discussed in the paper.
		
	\end{abstract}
	
	\maketitle
	
	%\linenumbers
	
	\section{Introduction}\label{sec:section1}
	
	The Coherent Elastic Neutrino-Nucleus Scattering (CE$\nu$NS) has been a productive research field since its first observation on CsI[Na] by COHERENT~\cite{COHERENT:2017ipa}. Low-energy neutrinos coherently scatter off the atomic nucleus as a whole, resulting in an enhancement to the cross-section, several tens to hundred times larger than the famous Inverse Beta Decay~(IBD) reaction used in the antineutrino detection. As a weak neutral current, the CE$\nu$NS becomes a new tool to study the neutrino properties and nuclear physics, from the improved bounds on non-standard neutrino interactions (NSI)~\cite{Dent:2016wcr,Liao:2017uzy,Dent:2017mpr,Farzan:2018gtr,Abdullah:2018ykz,Giunti:2019xpr}, to the constraints on the neutrino electromagnetic properties~\cite{Papoulias:2017qdn,Billard:2018jnl,Cadeddu:2018dux} and the weak mixing angle~\cite{Canas:2018rng,Cadeddu:2018izq,Huang:2019ene}, and to the nuclear structure~\cite{Cadeddu:2017etk,Ciuffoli:2018qem,Papoulias:2019lfi}.
	
	The more than 40 years gap from the proposal of CE$\nu$NS~\cite{Freedman:1973yd} to its first discovery~\cite{COHERENT:2017ipa} primarily owed to the tiny momentum transfer $Q$ between the neutrino and the nucleus. Although the cross-section is proportional to the square of the neutron number of the nucleus and much larger than the IBD and \textcolor{red}{$\nu$-e} elastic scattering processes, to satisfy the coherent condition, $Q$ should be smaller than $1/R$ where $R$ is the radius of the nucleus. This requires the neutrino energy to be smaller than $\sim$50~MeV and $Q$ is primarily concentrated in the range of \textcolor{red}{sub-keV/c to tens of keV/c}. The tiny momentum transfer is the only observable of the CE$\nu$NS process and puts stringent requirements on detectors, particularly on low energy detection.
	
	Motivated by the detection of the CE$\nu$NS process and the weak-interaction massive particles~(WIMP), the low-threshold and low background detection techniques are quickly developing~\cite{Klein:2022lrf}. The detection threshold is reaching 1~keV of nuclear recoil~(keV$_{\rm nr}$) in many detectors, such as cryogenic germanium detectors~\cite{CONUS:2020skt,Colaresi:2021kus,Wong:2015kgl}, liquid Noble gas detectors~\cite{DarkSide:2021bnz,Akimov:2022xvr,PandaX:2022aac,XENONCollaborationSS:2021sgk,LZ:2021xov}, CCD sensors~\cite{CONNIE:2019swq}. Although the inorganic crystal detector CsI[Na] was used in the first observation of CE$\nu$NS, the detection threshold is difficult to reduce to 1 keV$_{\rm nr}$ due to the smaller light yield than other techniques. Representative R\&D efforts on low-threshold inorganic crystal detectors are found in Refs.~\cite{NEON:2022hbk,Ding:2022jjm,Liu:2022cyy}.
	
	In this paper, we describe the development of a kg-scale cryogenic pure-CsI crystal detector running at 77 Kelvin. Compared to the previous work in Ref.~\cite{Zhang:2016szh}, scintillation photons were readout by two Hamamatsu S14161-6050HS-04 8$\times$8 SiPM arrays using cryogenic front-end electronics developed in Ref.~\cite{Wang:2021SiPM, Wang:2022zsv}. The wavelength shifter 1,1,4,4-Tetraphenyl-1,3-butadiene (TPB)~\cite{Burton:1973tla} was coated to shift the 340 nm scintillation light from CsI~\cite{2003NIMPA.504..307M} to around 420 nm where the detection efficiency of SiPM reaches the maximum. A world-leading light yield, 30.1 photo electrons~(p.e.) per keV energy deposit for X-rays and $\gamma$-rays with energies from 5.9~keV to 60~keV~(keV$_{\rm ee}$), was obtained. Considering a typical quenching factor of 6\% and the unprecedented light yield, a 4 p.e. detection threshold, which has been realized in other experiments~\cite{COHERENT:2017ipa,NEON:2022hbk}, corresponds to an energy threshold of 1~keV$_{\rm nr}$. This opens the door to using pure-CsI crystal in the detection of CE$\nu$NS of reactor antineutrinos.
	
	\section{The bench-top experiment}
	\label{experiment}
	
	There are several constraints to reaching the 1~keV$_{\rm nr}$ threshold in inorganic crystals. The first one is a finite scintillation light yield, usually $\sim$10~p.e./keV$_{\rm ee}$, corresponding to $\sim$1~p.e./keV$_{\rm nr}$. The triggered event will be dominated by the dark noise of the photo sensor if lower the trigger threshold to 2~p.e.. In addition, Cherenkov light produced by charged particles passing through the quartz window of a PMT also contributes much of the instrumental background with an energy less than 5~p.e.. Thus, a realistic trigger threshold is about 5~p.e., which corresponds to energy vary from 2 to 5~keV$_{\rm nr}$~\cite{COHERENT:2017ipa,NEON:2022hbk}.
	
	A promising way to overcome these limitations is to operate undoped CsI crystals with SiPM readout at low temperatures. As reported in many references and well summarized in Ref.~\cite{Klein:2022lrf}, the cryogenic undoped CsI/NaI crystal features a twice higher light yield than the doped crystal at room temperature. Operating SiPMs at low temperatures reduces the dark noises by several orders of magnitude to a similar level with dynode PMTs but provides a 100\% enhancement on the peak photon detection efficiency (PDE) than PMTs. There have been many R\&D efforts in this field~\cite{Zhang:2016szh,Wang:2022zsv, Liu:2022cyy,Lewis:2021cjv,Ding:2020uxu,Ding:2022jjm}, which indicates that it is an attractive approach for dark matter and neutrino detection using inorganic crystal detectors.
	
	Operating SiPM arrays and their front-end electronics at low temperatures is not trivial, particularly for the liquid nitrogen temperature, 77~K. The low-temperature performance of operational amplifier chips, capacitors, and resistors must be carefully considered. The light yield of pure-CsI peaks at this temperature~\cite{Amsler:2002sz} and the usage of both PMTs and SiPMs are being mature. Cryogenic front-end electronics is a must for SiPM arrays otherwise the single p.e.~signals would be overwhelmed by electronics noises. Benefiting from the recent developments~\cite{ Wang:2021SiPM, Wang:2022zsv}, a bench-top experiment has been set up with cryogenic SiPM arrays and front-end electronics. 
	
\subsection{Detector assembly}

	\begin{figure}[htb]
		\centering
		\includegraphics[width=8.5cm]{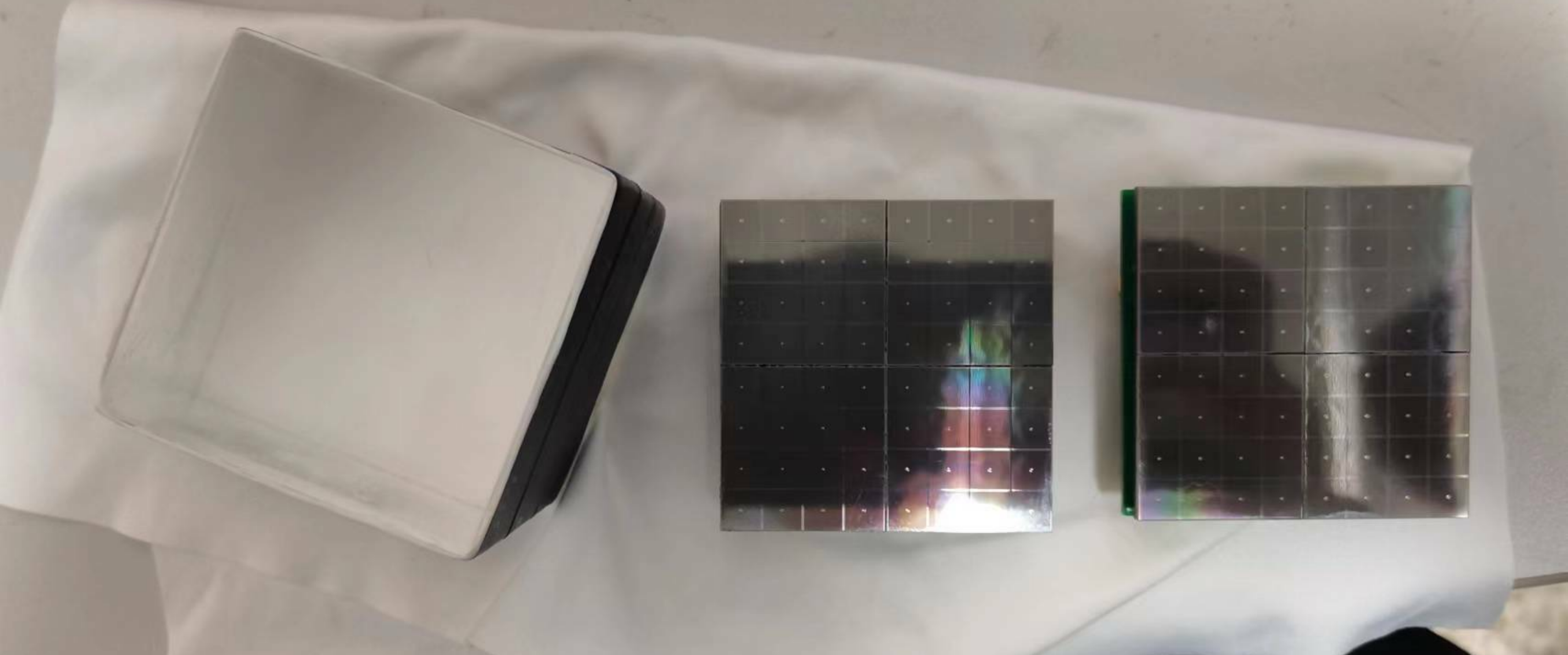}
		\caption{\label{Crystal_graph}Main detector components: one 2$\times$2$\times$2~inch$^3$ cubic pure CsI crystal (left) and two Hamamatsu S14161-6050HS-4 8$\times$8 SiPM arrays (middle and right). } 	
	\end{figure}
	
\begin{figure}[htb]
		\centering
		\includegraphics[width=5cm, angle=270]{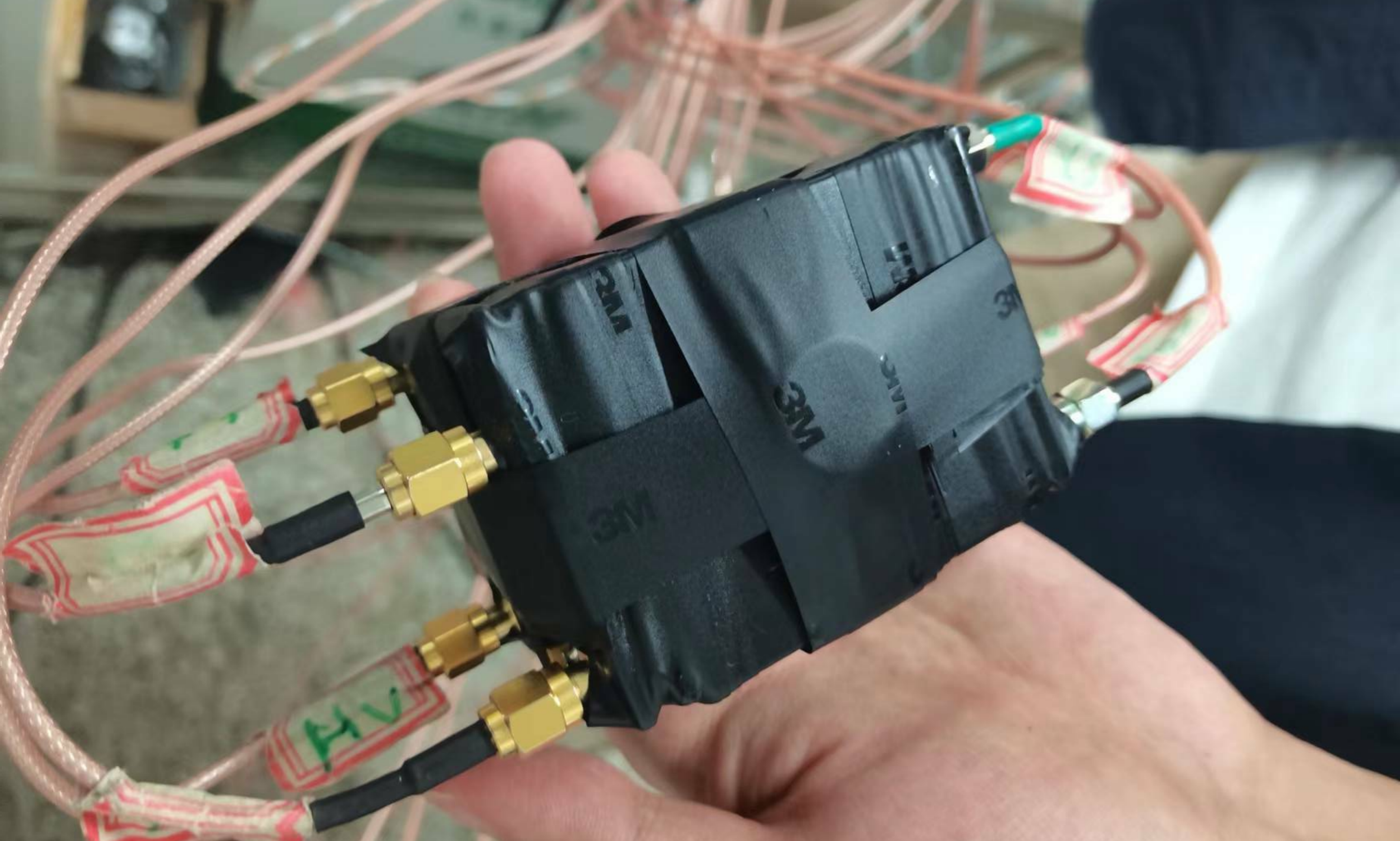}
		\caption{\label{detector_graph} The detector after assembly. Two SiPM arrays were opposite each other and combined with the crystal by silicone oil on the contact surfaces. The circular hump under the tape is a $\gamma$ source for calibration. } 	
	\end{figure}

	% paragraph of detector assembly
    The detector mainly consisted of a 0.56~kg pure CsI crystal with a size of 2$\times$2$\times$2~inch$^3$, and two Hamamatsu S14161-6050HS-04 8$\times$8 SiPM arrays with a size of 2$\times$2~inch$^2$, as shown in Fig.~\ref{Crystal_graph}. SiPM arrays were clamped on the opposite two sides while the other four sides were wrapped by Teflon to enhance the light collection. Silicone oil was used to provide better optical contacts between SiPM arrays and the crystal. Finally, sticky black tapes were used to stabilize the whole construction and operation in the cryogenic environment, as shown in Fig.~$\ref{detector_graph}$. 
	
	% paragraph of TPB
	The typical photon wavelength generated from a pure CsI crystal peaked at 310~nm at room temperature and shifted to 340~nm at 77~K~\cite{Woody:1990hq}. The Hamamatsu S14161-6050HS SiPM datasheet claims a photon detection efficiency (PDE) of about 25$\%$ at 340~nm wavelength at room temperature. To increase the wavelength matching, a wavelength shifter TPB was coated on the surface of SiPM arrays with a thickness of $\sim$150~$\mu$g/cm$^2$. TPB converts photons from 340~nm to 420~nm with almost 100\% efficiency.	PDE of SiPM arrays could reach 50$\%$ at 420~nm, which is significantly higher than all PMTs. 
	
\begin{figure}[htb]
		\centering
		\includegraphics[width=8.5cm]{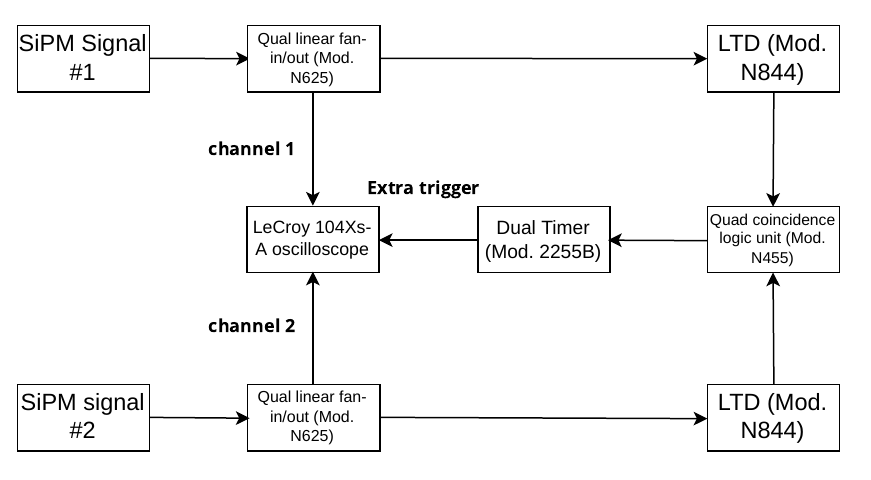}
		\caption{\label{Flow} The diagram of data flow. } 	
	\end{figure}

	% cryogenic system
	The cryogenic system was developed based on our previous experiments measuring characteristics of SiPMs from different vendors~\cite{Liu:2022cyy,Wang:2022zsv}.	Liquid nitrogen was used to cool the temperature of the whole system to 77~Kevin. Details about the cryogenic system can be found in the paper mentioned above. 
	% Electronics and DAQ
	The pre-amp circuit board adopted the same size design as the 2-inch SiPM array and was connected to the SiPM array through a multi-pin connector, which was compact and easy to arrange. The signals of the 2-inch SiPM array were connected in parallel to form a single-channel output signal after pre-amplification. 
    The power supply for the preamplifier and the bias voltage for the SiPM array were provided by RAGOL DP831A  and DH 1765-4 DC  power supplies respectively. A LeCroy 104Xs-A oscilloscope was used for data acquisition. The trigger signal for data acquisition adopted the method of coincidence of two SiPM arrays. \textcolor{red}{CAEN} modules were employed for signal splitting, discriminating, and coincidence between the two SiPM arrays. Figure~\ref{Flow} shows the data flow of the experiment.
    
    Once the oscilloscope received a trigger which was from the coincidence of two SiPM arrays, a 20~$\mu$s readout window would be opened with a sampling rate of 500~MS/s. The single channel threshold was set to 6~mV, corresponding to about 2$\sim$3~p.e.. The coincidence window of the quad coincidence logic unit was set to a width of 100~ns. $^{241}$Am and $^{55}$Fe sources were used to excite the crystal and were placed in the center of one surface of the crystal as shown in Fig.~\ref{detector_graph}. After the detector was built, we carried out several rounds of data acquisition at the temperature of liquid nitrogen. The experimental results are described in detail below.

	\subsection{Detector calibration results}
	\label{sec:section4}
\begin{figure}[htb]
		\centering
		\includegraphics[width=8.5cm]{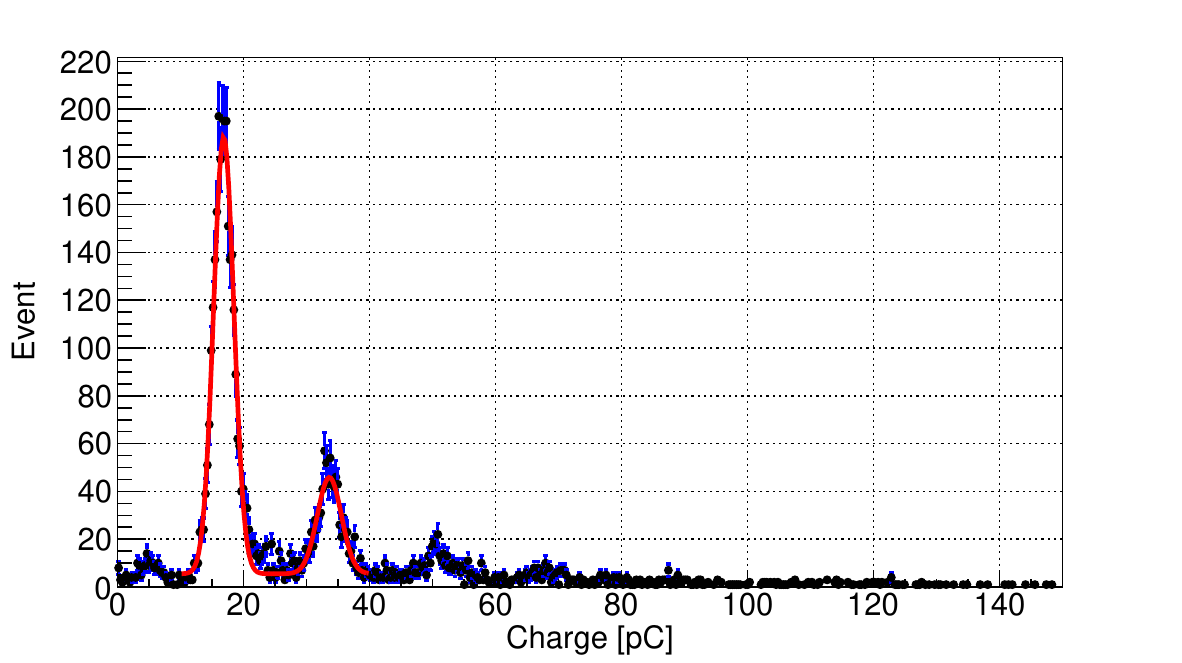}
		\caption{\label{SPEspectrum} Distribution of p.e. from one SiPM array at 37V.} 	
	\end{figure}

\begin{figure}[htb]
		\centering
		\includegraphics[width=8.5cm]{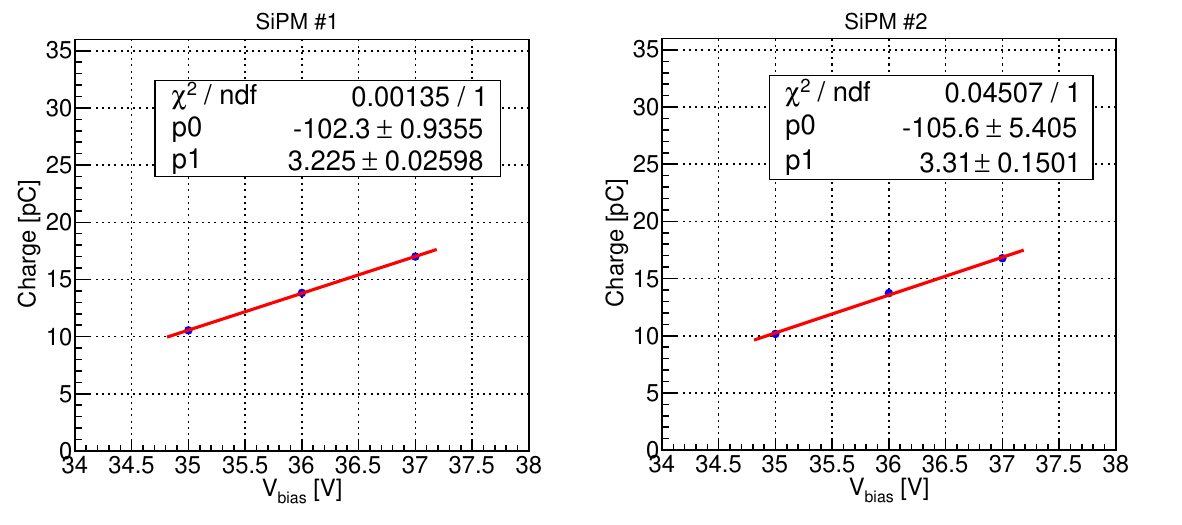}
		\caption{\label{SPE} Calibration results of SPE from two SiPM arrays.} 	
	\end{figure}

	The data analysis started from a single p.e. calibration of SiPMs. It was performed before measuring $\gamma$ spectra. At this time, the form of single-channel triggering was adopted, and the threshold was set to 0.5 p.e.. Figure~\ref{SPEspectrum} shows an example of the p.e.~spectrum while the bias voltage of two SiPM arrays was set to 37.0 V. The charge of a single p.e. under a certain voltage could be calculated according to the difference between 2 p.e. and 1 p.e. peaks. Both peaks were fitted by Gaussian functions simultaneously.	Figure~\ref{SPE} is the calibration results of two SiPM arrays. The SiPM arrays were running at three bias voltages to take the relationship between the PDE and bias voltages into consideration. The proportionality between voltages and SiPM gains is observed in Fig.~\ref{SPE}. Two red lines are plotted to evaluate the zero-gain point "V${_{breakdown}}$", which is literally a specific voltage where the gain of the SiPM array drops down to zero and is defined as the X-intercept of the fitting function. Usually, V${_{breakdown}}$ is slightly different for various SiPM arrays, although they are made by the same vendor.	In this experiment, V${_{breakdown}}$ of two SiPM arrays are 31.72~V and 31.90~V separately.

\begin{figure}[htb]
		\centering
		\includegraphics[width=8.5cm]{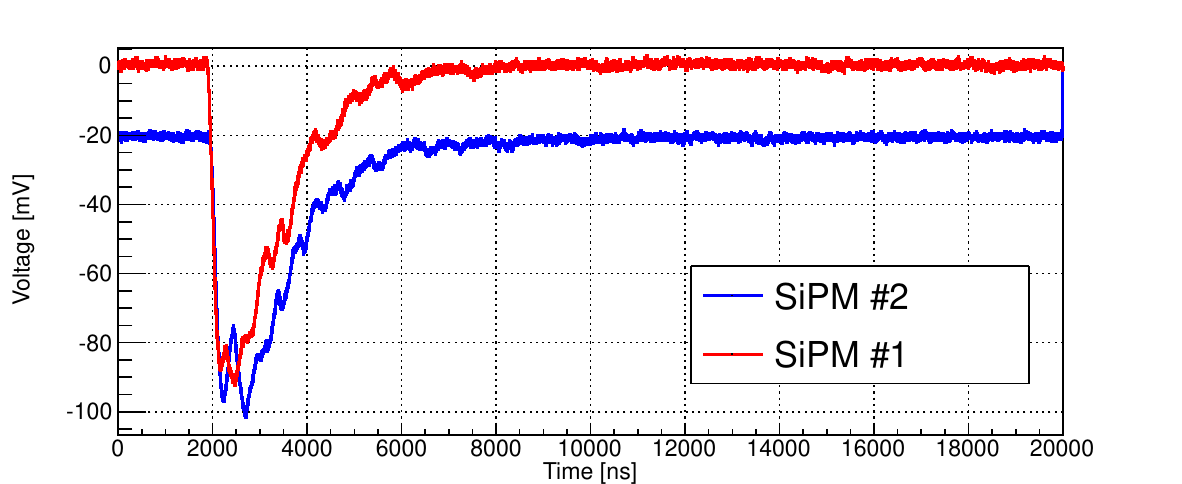}
		\caption{\label{example_pulse} An example of $\gamma$-ray output by two SiPM arrays.} 	
	\end{figure}

	Figure~$\ref{example_pulse}$ is an example of $\gamma$-ray pulses received by two SiPM arrays. The trigger was set at 2000~ns and the threshold was 10~mV. The baselines of two waveforms were manually shifted by 20~mV for better visibility.	The electronics noises were controlled below 2~mV which benefits from the good performance of the preamp circuit. 

\begin{figure*}[htb]
		\centering
		\includegraphics[width=15.5cm]{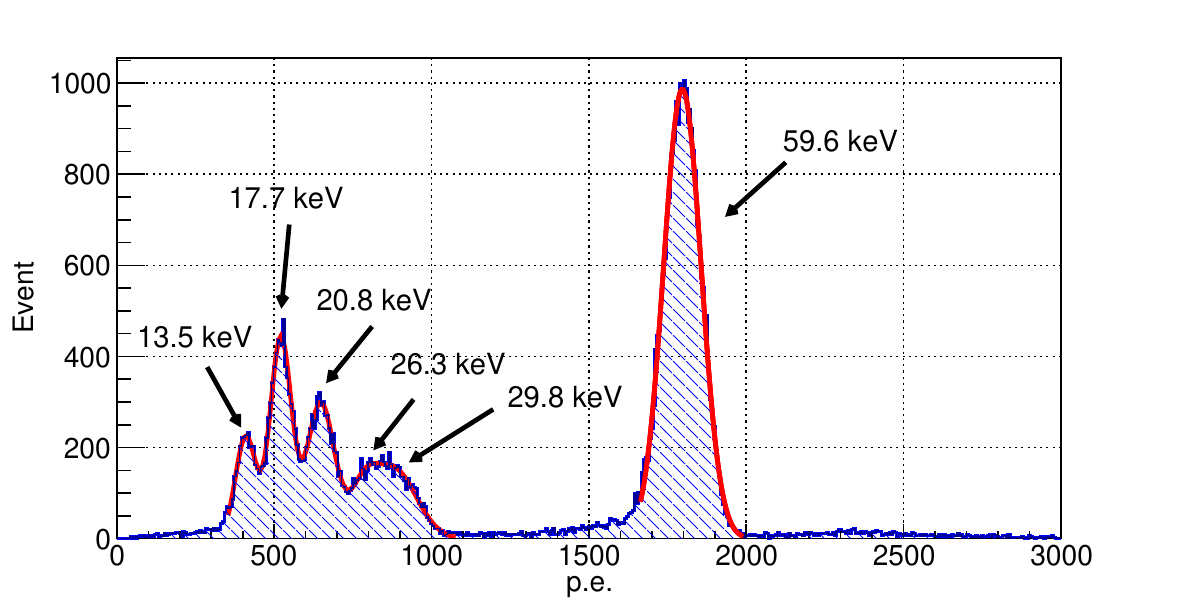}
		\caption{\label{Am241_spectrum} The measured $^{241}$Am spectrum. Six peaks are marked. Details are found in the text.} 	
	\end{figure*}

\begin{figure}[htb]
		\centering
		\includegraphics[width=9cm]{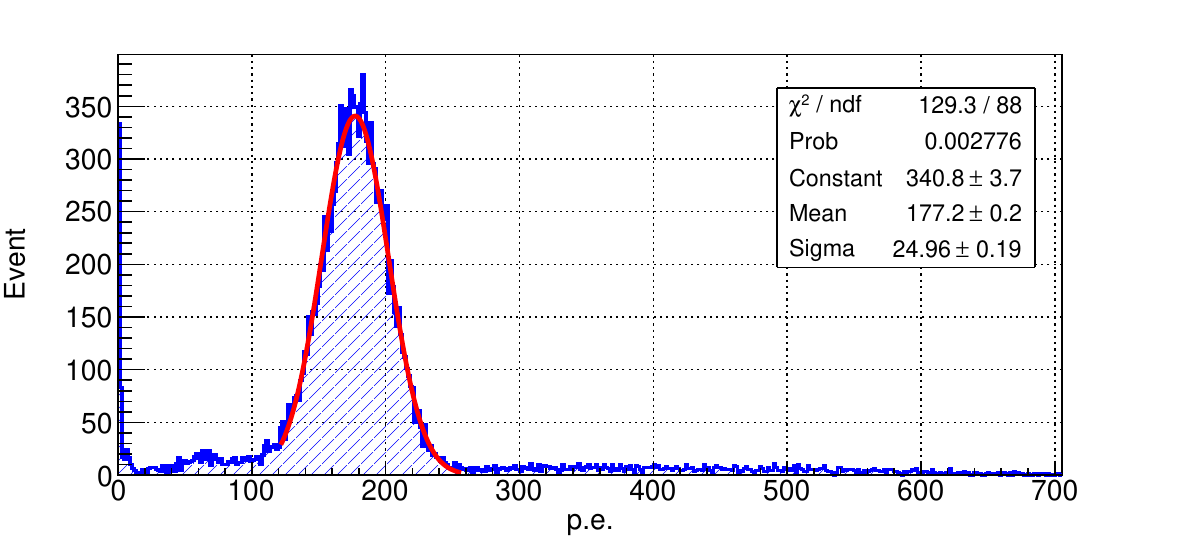}
		\caption{\label{Fe55_spectrum} Energy spectrum of a $^{55}$Fe source. The energy resolution of the 5.9~keV peak is about 14.1\%.} 
	\end{figure}

	After integrating the waveform and converting the charge to the number of photoelectrons,  Fig.~\ref{Am241_spectrum} shows the energy spectrum from a $^{241}$Am calibration source. To avoid amplifier saturation, the SiPM arrays' bias voltage was set to 34.0~V. The magnitude of the single photoelectron under this bias voltage was too small to be directly measured, and its charge was obtained by linear extension of the measured value in Figure 5.	Five distinct peaks have been marked out with arrows. The 59.6~keV peak was fitted with a Gaussian function. The other five peaks from 13.9~keV to 31.6~keV were jointly fitted with a sum of five Gaussian functions. The peaks at 13.9~keV, 17.7~keV, and 20.8~keV are attributed to the X-rays emitted by $^{241}$Am, the peak at 59.6~keV is caused by the $\gamma$-rays of $^{241}$Am, while the peak at $\sim$800~p.e. is composed of several distinct peaks. One is the 26.3 keV peak from the $\gamma$-rays of $^{214}$Am. The others are I-escape peaks of the 59.6 keV $\gamma$-rays. In the event of a $\gamma$-photon undergoing the photoelectric effect in a crystal, it creates a vacancy in the corresponding atomic shell layer, leading to the emission of X-rays or Auger electrons as outer electrons fill in the vacancy. The characteristic X-rays of the K-layer of the I atom have four energies, 28.6~keV, 28.3~keV, 32.3~keV, and 33.0~keV. If the photoelectric effect occurs near the surface of the crystal, these X-rays may escape from the crystal, resulting in an observed energy less than the energy of the incident photon.  Based on the branching ratios, the average energy of I K-shell X-rays is 29.8~keV. Therefore, the energy peak at $\sim$800~p.e. can be fitted with two Gaussian functions, one having a mean value of 26.3~keV and the other having a mean value of 29.8~keV. The 29.8~keV energy peak actually results from the combined effect of four Gaussian functions. Consequently, the energy resolution calculated in this way will be slightly degraded ~\cite{Tang,XRAY,xra}. Fitting results are listed in Table~\ref{EnergyResolution}.
	%	%
	The energy resolution is worse than that of photoelectron statistics, and one of the main contributing factors is optical crosstalk and afterpulsing. They not only affect energy resolution but also have a significant impact on the final light yield analysis. To obtain a realistic light yield, contributions from CrossTalk (CT) and AfterPulse (AP) should be estimated and subtracted carefully, which will be discussed in the following section together with the light yields provided. 
	To obtain the detector response at a lower energy region, a $^{55}$Fe source was also used. The measured spectrum of 5.9~keV X-rays is shown in Fig.~$\ref{Fe55_spectrum}$, and the peak is fitted with a single Gaussian function. Results are also provided in Table~\ref{EnergyResolution}.

	\begin{table}[htb]
		\centering
		\normalsize
		\begin{tabular}{cccc}
		\hline
		Energy~(keV) &  Resolution\\ \hline
		5.9 &  14.1\% \\
		13.9 &    7.9\% \\
		17.7 &    6.5\%\\
		20.8 &    6.3\%\\
		26.3 &    5.5\%\\
        29.8 &    8.2\%\\
		59.6 &   3.3\%\\
		\hline 
		\end{tabular}
		\caption{\label{EnergyResolution} The fitting results of the Am$^{241}$ source and the Fe$^{55}$ source. Resolution is defined by $\sigma$/mean of the Gaussian fits.}
	\end{table}

\begin{figure}[htb]
	\centering
	\includegraphics[width=9.5cm]{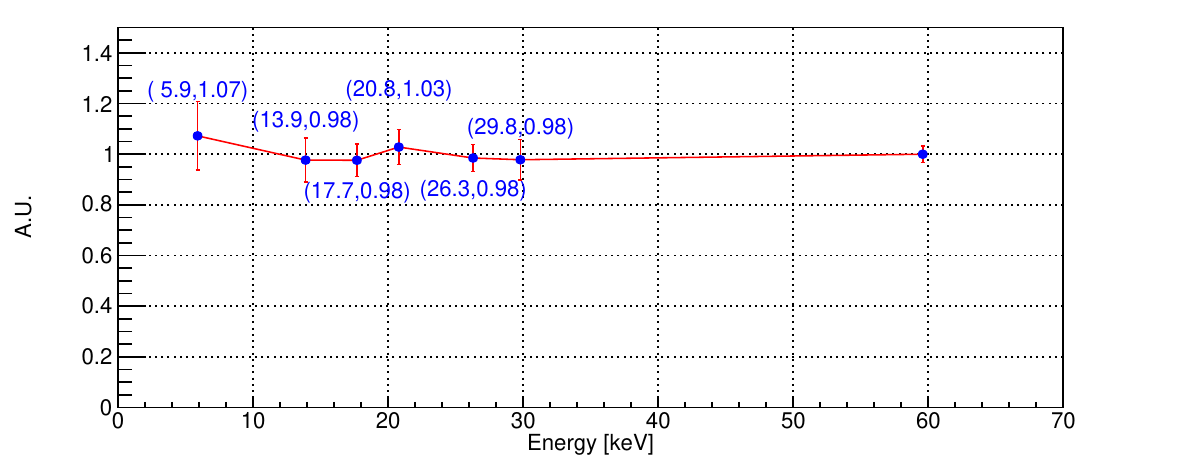}
	\caption{\label{EnergyLinearity} Energy linearity of the detector at V$_{bias}$=34.0~V. It is defined by the ratio of the light yield to the light yield at 59.6 keV. The highest point is 9.2$\%$ higher than the lowest point. Statistical uncertainties of these points are less than 1\%. The vertical error bars are used to represent the resolution of these peaks. } 	
	\end{figure}

	Six peaks above were used to represent the linearity of the crystal energy response in Fig.~$\ref{EnergyLinearity}$. The vertical error bars represent the resolutions of these peaks. These results exhibit good linearity, especially with no drop at energies below 10 keV. This performance is particularly important for measuring low-energy signals at the keV level and is significantly better than most doped crystals~\cite{AlekhinM.S.2013Ioe,AkimovD.2022Mosr,AngloherG.2017Rftf}.

\section{experimental results analysis}
	\label{sec:analysis}

%The contributions from dark noises accidentally falling into the signal window should be estimated and removed.	

	The measured light yields are affected by the bias voltage of the SiPM arrays and the usage of wavelength shifter TPB. From the instrumental point of view, larger bias voltages on the SiPMs indicated larger PDE and large CT and AP probabilities which ultimately result in a larger light yield. From the perspective of photon wavelength and SiPM response, the use of TPB can convert the light emitted by the crystal to a wavelength of 420~nm, which is more compatible with SiPM. This can significantly enhance the light yield. In this section, we will discuss these effects and provide the corrected light yields at various bias voltages with crosstalk and afterpulsing subtracted.

\subsection{Estimate of contributions from CT and AP}

The optical crosstalk of SiPM is mainly caused by the propagation of electron avalanche luminescence to adjacent pixels, causing the electron avalanche to occur in other pixels almost at the same time. More specifically, CT could be divided into internal CrossTalk (iCT) and external CrossTalk (eCT) according to the second fire position. For iCT, the second fire is just nearby the primary electron avalanche. But sometimes, the luminescence due to primary electron avalanches could escape from the SiPM and fire other SiPM arrays after traveling a distance when multiple SiPM arrays exist in the detector. This is called eCT. At the micro level, eCT could be observed directly by a CCD (Charge-Coupled Device) camera in the dark environment~\cite{McLaughlin:SiPMeCT,Miroyan:SiPMeCT}. The other contribution that should be considered is AP, which is derived from the avalanche of a trapped electron in the same pixel. 

\begin{figure}[htb]
	\centering
	\includegraphics[width=8.5cm]{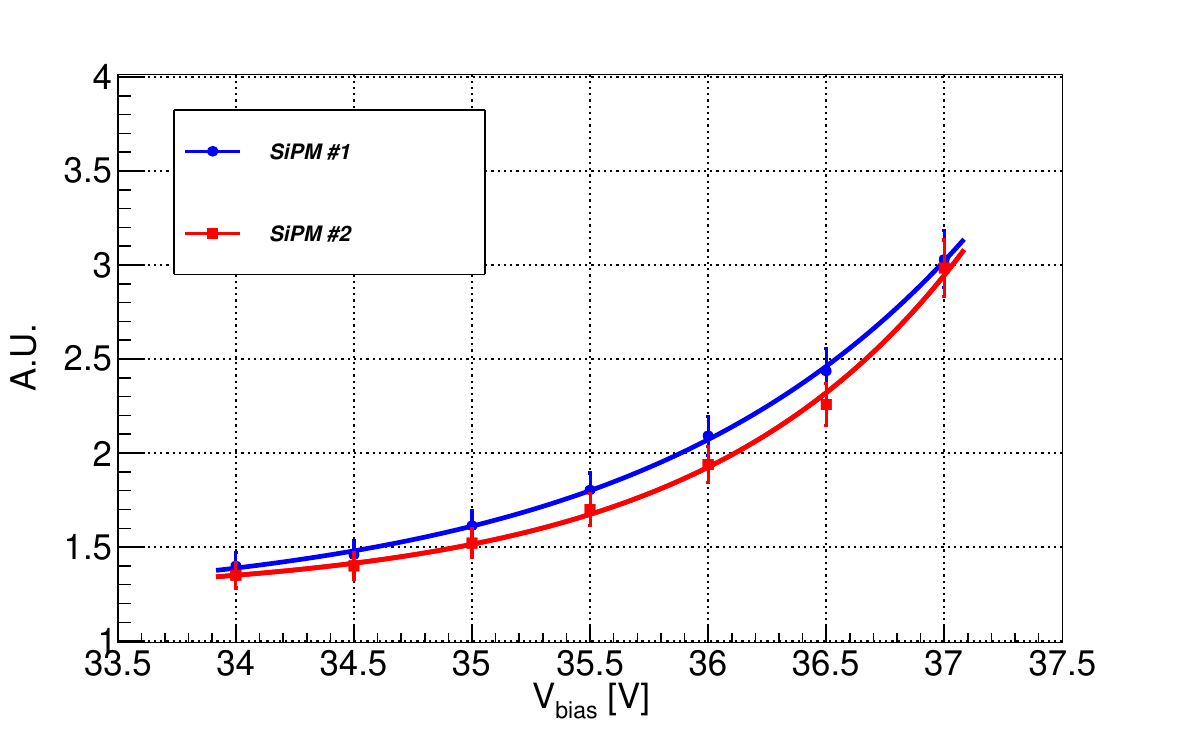}
	\caption{\label{iCT} Correction factor of iCT and AP with bias voltage from two SiPM arrays respectively.} 	
\end{figure}

Our previous works have presented a quantitative analysis in respect of iCT and AP for Hamamatsu S14161-605-HS SiPM array~\cite{Wang:2022zsv,Wang:2022LAr}. Specially, we given the exact ratio of iCT and AP in the dark environment~\cite{Wang:2022zsv}, and Ref~\cite{Wang:2022LAr} introduced a new method to calculate the overestimation factor of the light yield duo to the exist of iCT and AP. The main idea of this method is to simulate an energy distribution using dark signals (including iCT and AP) or using single photoelectron signals separately. The ratio of the two distributions' mean statistically represents the correction factor of real light yield. Based on the method, the correction factors corresponding to the two SiPM arrays' iCT and AP were estimated in Fig~$\ref{iCT}$ respectively. 

\begin{figure}[htb]
	\centering
	\includegraphics[width=8.5cm]{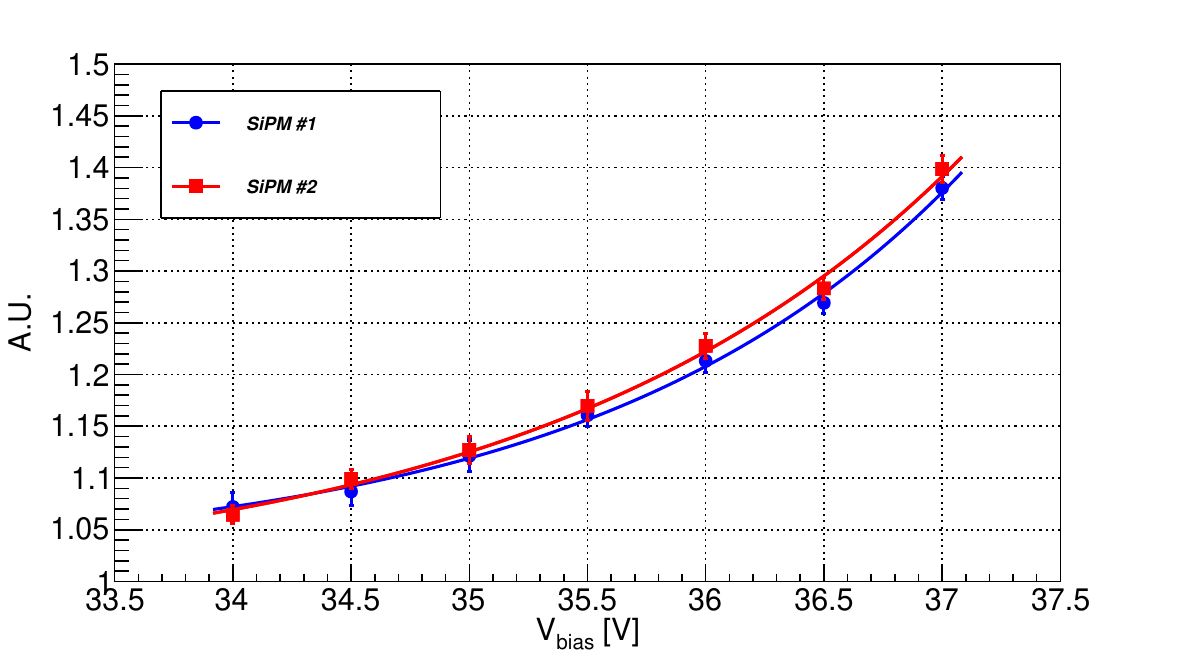}
	\caption{\label{eCT} Correction factor of eCT with bias voltage from two SiPM arrays respectively. The light yield is scaled to 0V.} 	
\end{figure}

Furthermore, eCT also could be measured accurately in the experiment for individual SiPM arrays. For instance, the paper~\cite{DarkSide:CT} introduced a simple way to estimate the performances of eCT in a certain detector. The eCT performances of two SiPM arrays in our detector were estimated based on the reference. That is, the light yield of one SiPM array was measured with the changing bias voltage of the opposite SiPM array. The light yield difference because of the bias voltage change of the opposite SiPM is exactly the result of eCT. The results are illustrated in Fig~$\ref{eCT}$.

Conclusively, the contributions of all three effects could be subtracted from the measured light yield of the detector. 

\subsection{Light yield}

\begin{figure}[htb]
		\centering
		\includegraphics[width=9cm]{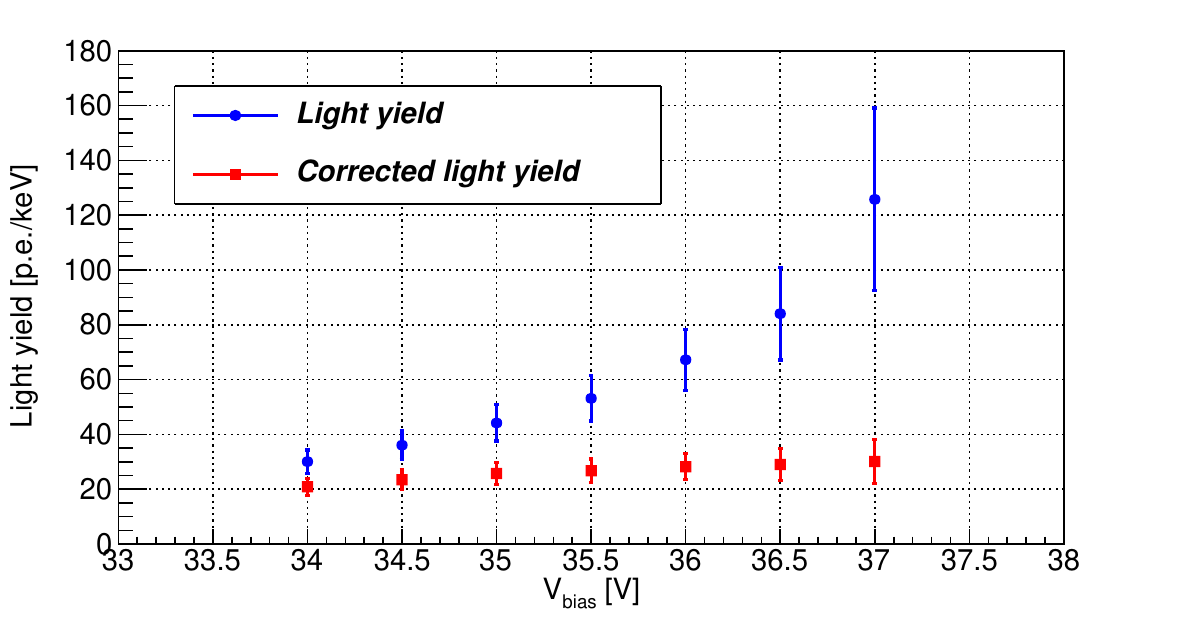}
		\caption{\label{LY} 
			The blue circle points represent the light yield of the pure CsI crystal with two SiPM arrays. As expected, the enhancement of light yield with increasing V$_{bias}$ could be observed. The light yield at V$_{bias}$ = 37~V reaches up to 122.7 p.e./keV. In contrast, the estimated light yield after correcting for CT and AP was drawn on the red dotted line. More details were included in Table~\ref{LYtable}. } 	
	\end{figure}

	As one of the most critical parameters for a low threshold detector, we have measured a group of light yields at different bias voltages, from 34.0~V to 37.0~V with a step of 0.5~V using two different radioactive sources. In general, with the bias voltage increases, both the Photon Detection Efficiency (PDE) and the CT and AP probabilities increase. The measured light yield results are shown in Fig.~$\ref{LY}$ and Table~\ref{LYtable}. A positive correlation between the bias voltages and the light yields is observed. After eliminating the effects of iCT, eCT, and AP, the corrected light yield of the 2-inch pure CsI crystal with SiPM arrays is finally shown in Fig.~$\ref{LY}$ as square points. A slow increase is still observed mainly due to the enhancement of PDE, which is the only factor that could change the detector's light yield with V$_{bias}$. The calibration of PDE at liquid nitrogen temperature will be carried out in our follow-up research. 
    
	\begin{table}[htb]
		\centering
		\footnotesize
		\begin{tabular}{ccccc}
			
			\hline
			V$_{bias}$ (V) & p.e. & resolution  & measured LY & corrected LY\\ \hline
			34.0 & 176.8&   14.3\%&		30.0$\pm$4.3 & 20.9$\pm$3.1\\
			34.5 & 206.4&	14.4\%&		35.0$\pm$5.0 & 23.4$\pm$3.6\\		
			35.0 & 241.4&	14.4\%&		40.9$\pm$5.9 & 25.8$\pm$4.0\\
			35.5 & 286.7&	15.4\%&		48.6$\pm$7.5 & 26.7$\pm$4.3\\
			36.0 & 359.4&	16.2\%&		60.9$\pm$9.9 & 28.2$\pm$4.8\\
			36.5 & 473.5&	19.7\%&		80.2$\pm$15.8 & 29.0$\pm$5.9\\
			37.0 & 723.8&	26.0\%&		122.7$\pm$32.2 & 30.1$\pm$8.1 \\ \hline 
		\end{tabular}
		\caption{\label{LYtable} Light yields of the pure CsI crystal measured using a $^{55}$Fe source. The second and the third columns are the mean p.e. and resolution of the 5.9~keV X-ray peak. The measured LY is p.e. per keV$_{\rm ee}$ including contributions from CT and AP, while the corrected LY is the light yields with CT and AP effects corrected. The error of the corrected LY consists of the statistical error from measured LY and correction factors.
		}
	\end{table}
	
\begin{figure}[htb]
		\centering
		\includegraphics[width=9cm]{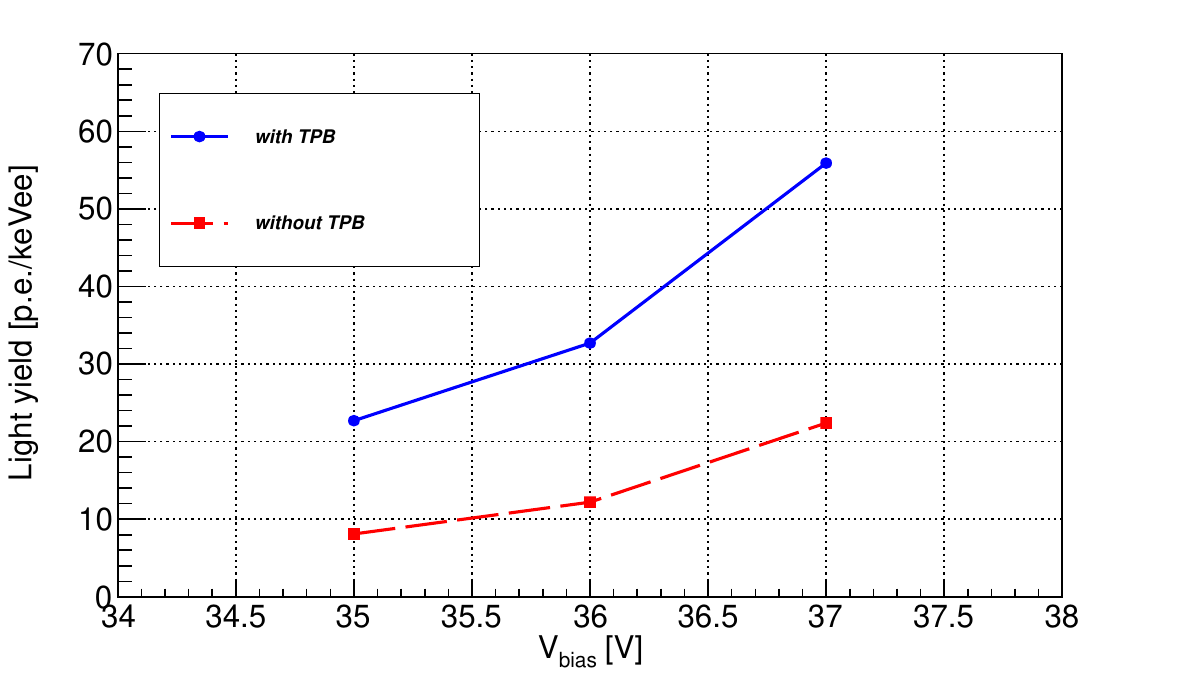}
		\caption{\label{TPBplot} Light yields of two SiPM arrays with and without TPB painting. } 	
	\end{figure}

	As mentioned, the surfaces of SiPM arrays were covered by TPB to improve the light yield. Theoretically, light yield could increase about $\sim$100$\%$ by using TPB, because QE of SiPM would increase from $\sim$25$\%$ at 340~nm to $\sim$50$\%$ at 420~nm. It is tested with an experiment using two SiPM arrays, but one array was not coated with TPB. Figure~\ref{TPBplot} shows the measured results without CT correction at three V$_{bias}$ values.  The red and blue lines represent the light yields measured with the SiPM array without and with TPB, respectively. The increase demonstrates the importance of using TPB.

\begin{figure}[htb]
	\centering
	\includegraphics[width=9cm]{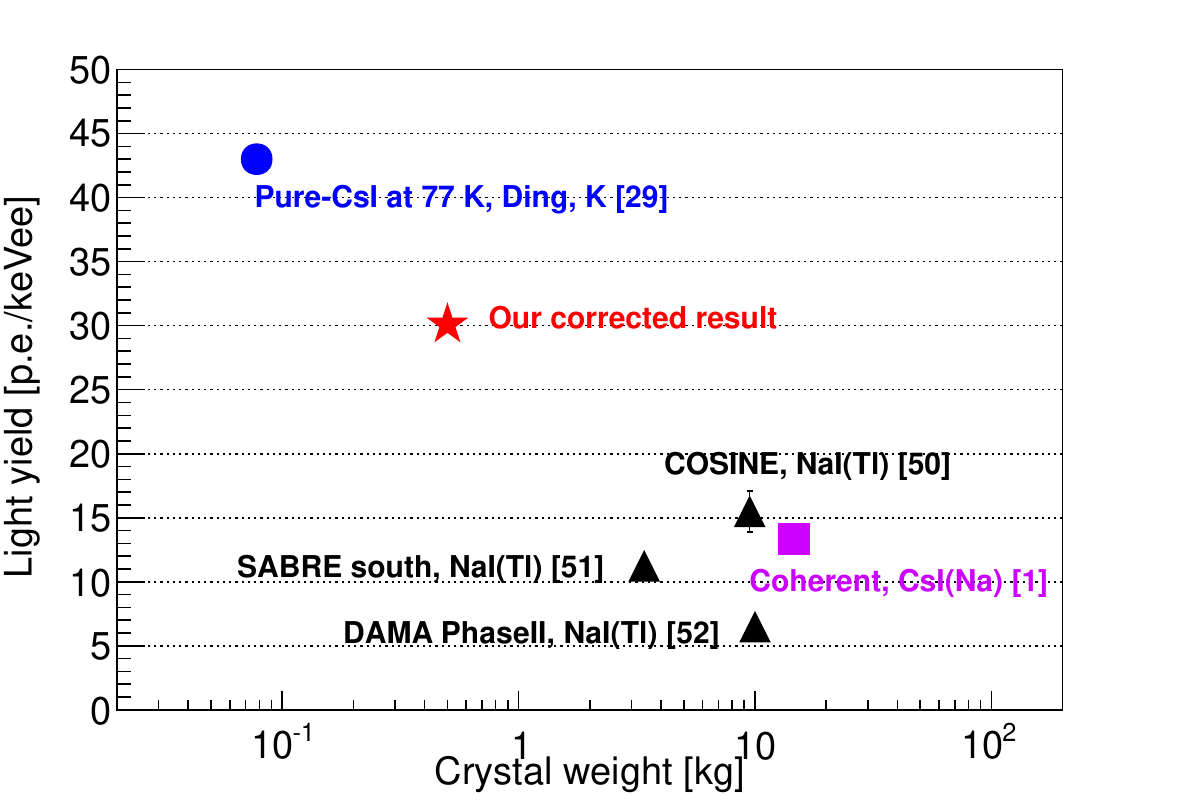}
	\caption{\label{LY_Graph} Global picture of light yields in typical inorganic crystals~\cite{COHERENT:2017ipa,Ding:2022jjm, COSINE-100, DAMA:2012, SABRE:2021}.} 	
	\end{figure}

	We compare our results with other crystal detectors in literature as shown in Fig.~$\ref{LY_Graph}$. Experiments using doped crystals featured light yields smaller than 20~p.e./keV$_{\rm ee}$. Recent results in Ref.~\cite{Ding:2022jjm} used a 0.6$\times$0.6$\times$1.0 cm$^3$ cryogenic pure CsI crystal readout by SiPM chips and achieved about 43~p.e./keV$_{\rm ee}$. It's not trivial to perform an apple-to-apple comparison between our results with it because the CT effect was not corrected in Ref.~\cite{Ding:2022jjm}.
	
	Conclusively, to our current knowledge, we achieved one of the highest light yields among all similar scintillator detectors. we measured a 30.1~p.e./keVee light yield of pure CsI crystal with SiPM readout at V$_{bias}$ = 37~V, corresponding to a V$_{over voltage} \approx$ 5.2~V. V$_{over voltage}$ is the difference between the supplying bias and the breakdown voltage.
	
	However, the energy resolution under this high voltage is not satisfactory enough because CT and AP events are dominating the signals with increasing voltage. In contrast, it is convinced that V$_{bias}$ = 35~V, corresponding to a V$_{over voltage} \approx$ 3.2~V, is a better choice since the detector shows a good resolution and a good light yield simultaneously. In the next section, the calculation of the CE$\nu$NS detection potentials would be based on the condition of V$_{bias}$ = 35~V.
	
\section{Potentials in the CE$\nu$NS detection}
\label{sec:physics}
	
	Detection of the CE$\nu$NS process requires low threshold detectors. For a scintillation detector, a larger light yield means a smaller detection threshold, which corresponds to a better sensitivity in the new physics searches.	There have been proposals of using cryogenic CsI detectors at spallation neutron sources~\cite{Baxter:2019mcx,Akimov:2022oyb} and exciting results are foreseen. This paper discusses the possibility of using cryogenic CsI for the reactor neutrino CE$\nu$NS study.
	
	In a commercial reactor, electron antineutrinos are produced from the fission products of four major isotopes, $^{235}$U, $^{238}$U, $^{239}$Pu, and $^{241}$Pu. We generate the reactor neutrino spectrum using the following method. The average fission fractions are assumed to be 0.58, 0.07, 0.30, 0.05, with mean energies per fission of 202.36 MeV, 205.99 MeV, 211.12 MeV, 214.26 MeV~\cite{Ma:2012bm} for the above four isotopes, respectively. The neutrino energy spectra per fission of $^{235}$U, $^{239}$Pu, and $^{241}$Pu are taken from Huber~\cite{Huber:2011wv}, and of $^{238}$U is taken from Mueller~\cite{Mueller:2011nm}.  In the CE$\nu$NS detection of reactor neutrinos, the high energy part is of great importance. We apply an exponential extrapolation in the 8~MeV and 10.5~MeV energy region and cut off at 10.5~MeV according to the measurement by the Daya Bay experiment~\cite{DayaBay:2022eyy}. The extrapolated part is reduced by a factor of 0.7 according to measurement at Daya Bay~\cite{DayaBay:2022eyy}. Furthermore, discrepancies have been found between the data and the Huber-Mueller spectra, a 5\% flux deficit, and a spectral distortion in the 5 to 7 MeV region. Both deficits were corrected in our study.

\begin{figure*}[htb]
	\centering
	\includegraphics[width=15cm]{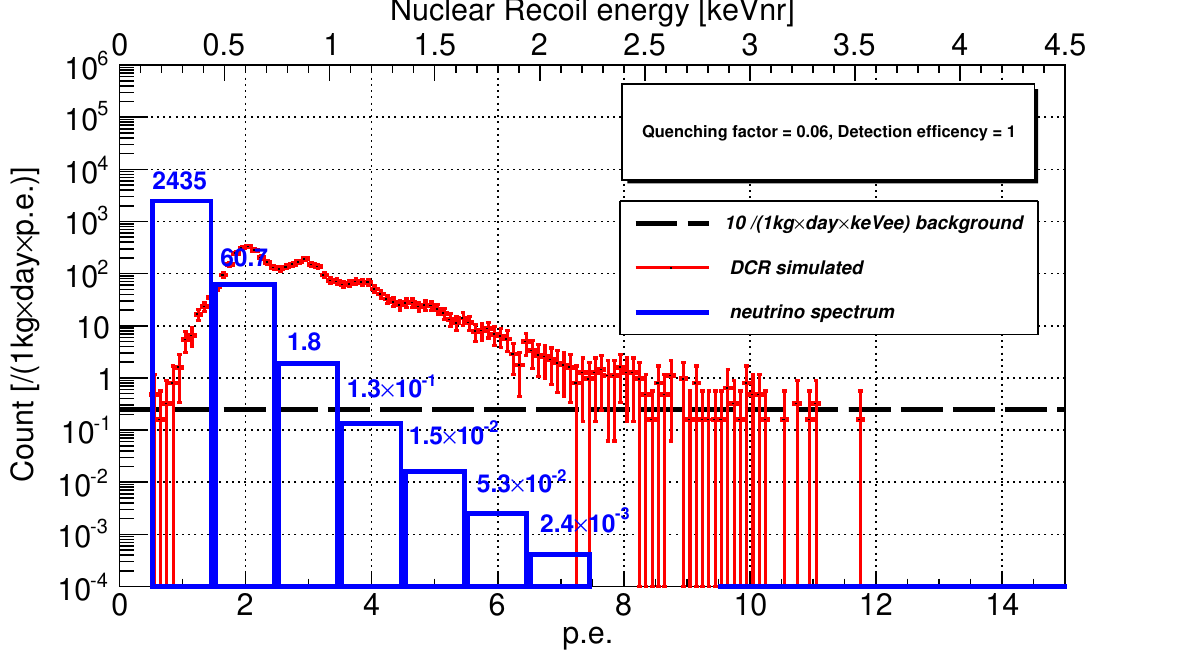}
	\caption{\label{Neutrino} Blue: neutrino p.e. spectra for the cryogenic pure CsI detector that is located 35 meters away from a 4.6~GW$_{\rm th}$ reactor. Numbers upon bins represent how many CE$\nu$NS events could be detected at this p.e. by setting the detection efficiency to 1 and the quenching factor to 0.08. Red points with error bar: MC simulation result of the DCR distribution of two 2-inch SiPM arrays with the 0.5 p.e. coincidence threshold. Black dotted line: an assumed crystal background level, corresponding to 10 events per kg per day per keVee. The X-axis is the total p.e. collected by the two SiPM arrays.}
	\end{figure*}

We put a 35~m distance between a 4.6~GW$_{\rm th}$ reactor.	The total flux of reactor neutrinos is about 6$\times10^{12}$/s/cm$^2$. The CE$\nu$NS cross-section uses the prediction of the Standard Model.	Setting the detection efficiency to 1 and the quenching factor to 0.06~\cite{Akimov:2022oyb}, the nuclear recoil spectrum is converted to the total p.e. distribution according to the light yield measured. As mentioned, the light yield used is 40.9~p.e./keV$_{\rm ee}$ at V$_{bias}$ = 35V. As shown in Fig.~$\ref{Neutrino}$, benefiting from the high light yield, the CE$\nu$NS signal events is around 2~p.e. is expected to be more than 60 per kg per day.

\begin{figure}[htb]
		\centering
		\includegraphics[width=8.5cm]{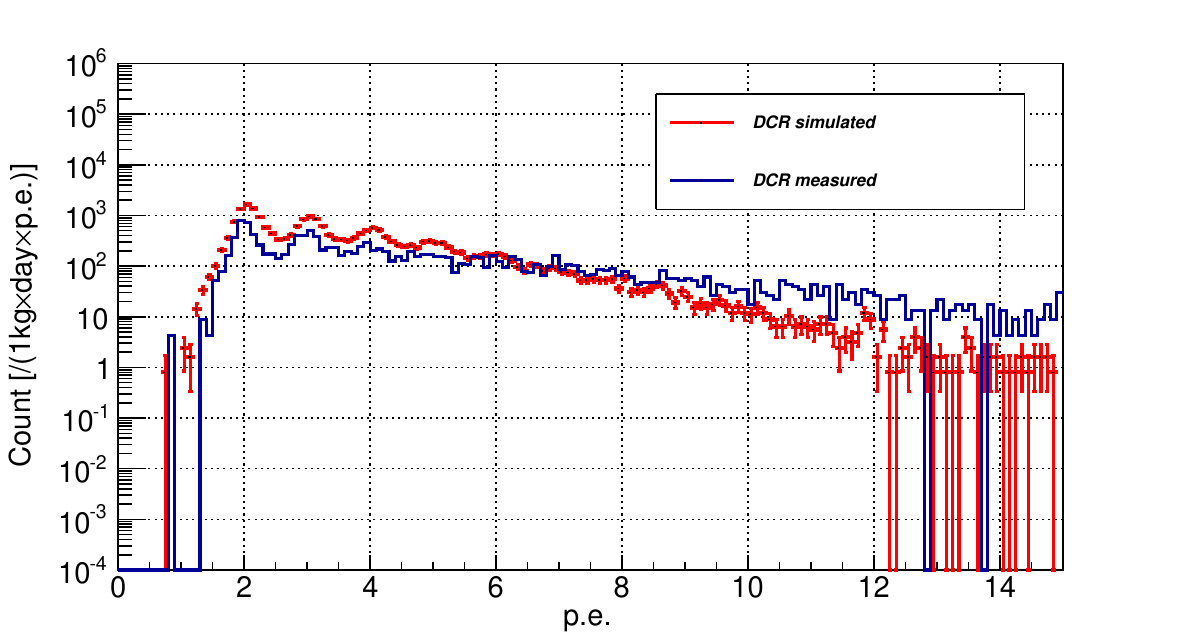}
		\caption{\label{DCR_Distribution} \textcolor{red}{The measured and simulated DCR distributions at V$_{bias}$ = 37~V.}} 	
	\end{figure}

    The most crucial background in this energy region is not from radioactivity, but the coincidence of SiPM dark counts. The dark count rate~(DCR) of Hamamatsu S14161-6050HS SiPM is $\sim$0.1~Hz/mm$^{2}$ with V$_{bias}$ of 35.0~V at liquid nitrogen temperature, corresponding to $\sim$280 Hz per 2-inch SiPM array. An MC simulation based on this DCR level was performed for the evaluation of the dark signal effect assuming that the coincidence thresholds of two SiPM arrays are set to 0.5~p.e. with the coincidence window of 100 ns. The principle of the MC simulation is described below. 
    The probability of finding a dark signal in a window of 1~ns was calculated by: 
	\begin{equation}
		P(1~ns)=280Hz\times1ns
		\label{P_1ns}
	\end{equation}
	which is about 2.8$\times$10$^{-7}$. As a result, we constructed two simulated SiPM output channels with a readout window length of 1~$\mu$s. In each nanosecond of the window, if a dark count was found generated according to $P(1~ns)$, a measured SiPM dark signal was added in this nanosecond. If within 100~ns, the other channel was also triggered by dark noises, and both channels passed $\sim$0.5 p.e.~threshold, the coincidence was recorded and the energy was calculated by integrating the waveform. The simulated result is shown in Fig.~$\ref{Neutrino}$ with the red dots and error bar.	The dark noise background is much higher than neutrino signals. \textcolor{red}{Besides, Fig.~\ref*{DCR_Distribution} shows the comparison between the simulated and measured DCR distributions at V$_{bias}$ = 37~V. The disparity observed in the high p.e. region is attributed to the presence of an increased number of large pulses in the measured data, which are caused by the environmental radioactivities and cosmic rays and were not considered in the simulation.}
		
	Another background is the radioactivity background contributions, from both the external materials and CsI crystal itself. A dedicated simulation and measurement are ongoing and here we just assume radioactivity background contributes 10 events per kg per day per keVee at such a low energy scale of around 100~eV.

	In the calculation of observing CE$\nu$NS, a specific threshold setting was used. That is, we set the threshold of both SiPM arrays to 0.5 p.e. to 1.5 p.e., which means the detector would be only sensitive to CE$\nu$NS events that exactly produce two photoelectrons. A benefit of this setup is the threshold between 0.5 p.e. to 1.5 p.e. could largely eliminate the effect of iCT since the iCT signal at least has an amplitude of 2 p.e. Another significant reason is that CE$\nu$NS events dominated in the energy region of 2 p.e. for our detector. If the trigger is enhanced to 2 p.e. for both SiPM arrays, there are only two CE$\nu$NS events per day as shown in the fourth bin of Fig.~$\ref{Neutrino}$ and the dark counts' coincidence dominate at the region. It is derived from the intrinsic characteristic of Hamamatsu S14161-6050HS-4 unless we find better alternatives that could work in liquid nitrogen with less dark noise. 

	The Confidence Level (C.L.) of observing CE$\nu$NS by operating a 1~kg pure-CsI detector was evaluated with operating time using the following formula:
	\begin{equation}
		C.L.=\frac{N_\nu}{\sqrt{\textcolor{red}{12}{\times}N_{bg}+N_{dcr}+N_\nu}}.
		\label{Eq.Q}
	\end{equation}

	\begin{figure}[htb]
		\centering
		\includegraphics[width=8.5cm]{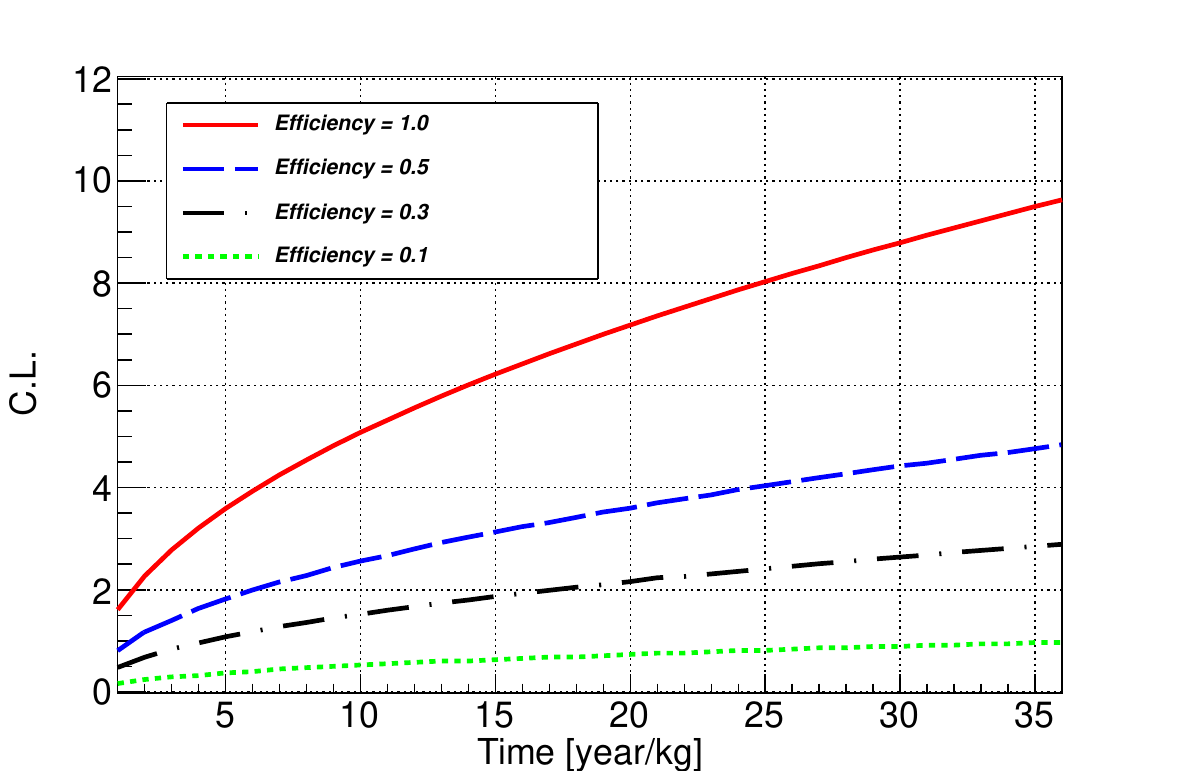}
		\caption{\label{CL} C.L. performance of a 1~kg pure-CsI detector with operation time. \textcolor{red}{When calculating the C.L., we assumed that the reactor is inactive for 1 month annually, during which background measurements are conducted and the rest time is utilized for signal measurements.}}

	\end{figure}

	In this equation, $N_\nu$ is the expected counting rate induced by neutrino, $N_{bg}$ and $N_{dcr}$ correspond to the radioactivity background counting rate and the dark noise counting rate. \textcolor{red}{The factor of 12 preceding N$_{bg}$ arises from the assumption that the reactor is inactive for one month annually, during which background measurements are conducted, while the rest are operational and utilized for signal measurements.} We also assume there are no reactor-related backgrounds. Firstly, the detector is located 35~m from the reactor core, which is predominantly shielded by concrete. Therefore, $\gamma$s or neutrons produced by the reactor can not reach our detector. Secondly, we conducted actual measurements of the neutron flux at the experimental site, which showed no significant difference from the regular environment. Then assuming that $N_{dcr}$ can be calculated precisely using the $in-situ$ monitoring of dark noise, and $N_{bg}$ can be extracted using reactor-off data. Thus, a coefficient 12 was added to $N_{bg}$ to consider this effect that there is one month for reactor-off data-taking every year. Fig.~\ref{CL} shows the expected sensitivity of observing CE$\nu$NS near a 4.6~GW$_{\rm th}$ using a 1~kg CsI detector with operating time. If the target mass of the detector can be increased to 20~kg, we hold the potential to observe the CE$\nu$NS process of reactor neutrinos through one year of operation.

\section{Discussion of external CT background}
	
\begin{figure*}[htb]
	\centering
	\includegraphics[width=15cm]{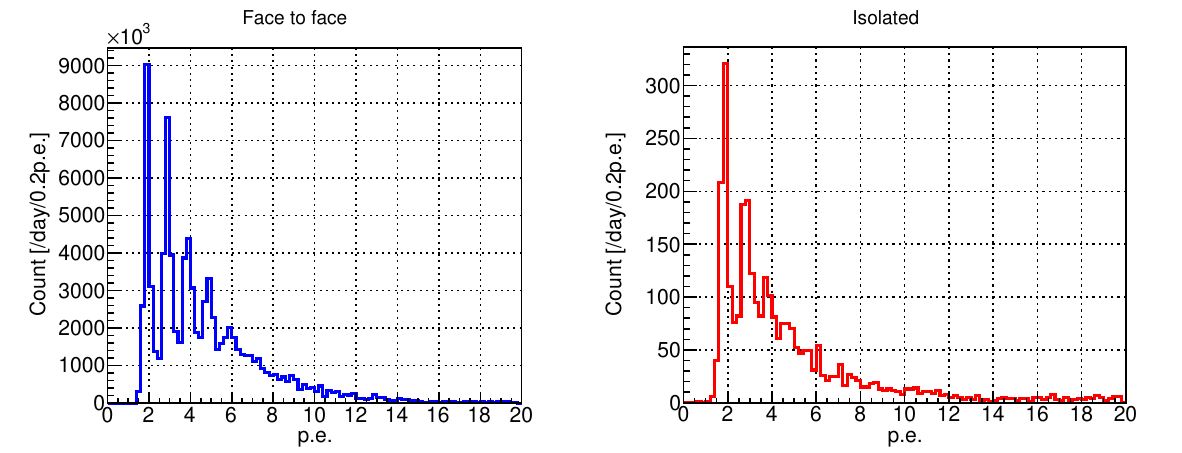}
	\caption{\label{eCTdiffernece} The coincidence distribution of two SiPM arrays at a dark environment under liquid nitrogen temperature. One day's data was included. The threshold of both SiPM arrays was still set to 0.5 p.e., and the coincidence window is 100~ns. The blue spectra left-side is the coincidence distribution of two SiPM arrays face-to-face closely without any block. So the effect of eCT gets maximum in the setup. In contrast, the red spectra right-side are the coincidence distribution of two SiPM arrays that are separated from each other by a black cloth. eCT photons from the surface of one SiPM array are no longer able to hit another SiPM. It represents the real coincidence count from dark noises.} 	
	\end{figure*}

	By optimizing the detector's threshold, we have shown that it is possible to use a pure-CsI detector with SiPM readout for reactor CE$\nu$NS detection, even though the recoil energy of reactor neutrinos is on the keV scale. At such low energy levels, the primary background would be accidental coincidence events caused by the dark noises of two SiPMs. However, other coincidence events would also be a serious background when SiPM's threshold drops down to a very low level. They are the combination of a SiPM dark noise and the coincident signal of eCT caused by the opposite SiPM. Fig.~$\ref{eCTdiffernece}$ shows the distribution of coincident events with two SiPM arrays. The overwhelming eCT coincidence background could be observed according to different experiment setups when dropping down the threshold of the detector.  
	
	In conclusion, the luminescence of a SiPM array cathode can result in a significantly high background counting rate in low threshold detectors. The eCT background count rate is approximately five orders of magnitude higher than the normal accidental coincidence dark count rate, even at liquid nitrogen temperature. If there is no further analysis or method to discriminate the eCT background from target events, all target events would be overwhelmed by this background noise. We are currently investigating the impact of SiPM eCT on low-threshold detectors, and the relevant results will be reported in our forthcoming articles.

\section{Conclusion}

	Low-threshold neutrino detectors are opening up new avenues in low-energy neutrino physics.	For detectors only sensitive to photons, increasing the light yield is most important to lower the energy threshold. In this paper, we have obtained the world-leading light yield in the CsI crystal by operating it at liquid nitrogen temperature, namely 30.1~p.e./keV$_{\rm ee}$ in a kilogram-scale pure CsI detector, much larger than doped crystals. This owes to the combination of cryogenic SiPM and front-end electronics, and the coating of wavelength shifter TPB. For this experiment, the current issue lies with the eCT of SiPM. If it is resolved, running a 20~kg detector at 35~m distance to a 4.6~GW$_{\rm th}$ reactor, we have a good chance to observe the reactor neutrino CE$\nu$NS process on Cs and I nuclei in one year.
	  
\section{Acknowledgement}

The study is supported in part by the National Natural Science Foundation of China (Grant No.~12275289 and Grant No.~11975257), the State Key Laboratory of Particle Detection and Electronics (SKLPDE-ZZ-202116), and the Youth Innovation Promotion Association of Chinese Academy of Sciences (2023015).

\bibliographystyle{apsrev4-1}
\bibliography{CsI}

\end{document}